\documentclass{elsart}

\usepackage{graphicx}
\usepackage{amssymb}

\begin{document}

\begin{frontmatter}

\title{Influence of deterministic trend on the estimated parameters of
  GARCH(1,1) model}

\author{C\u{a}lin Vamo\c{s} \thanksref{label2}}, 
\author{Maria Cr\u{a}ciun \corauthref{cor1}\thanksref{label3}}
\thanks[label2]{First author was supported by grant 2-CEx06-11-96/19.09.2006.}
\corauth[cor1]{Corresponding author. Email address:
craciun@ictp.acad.ro}

\thanks[label3]{Second author was supported by MEdC-ANCS under grant  ET 3233/17.10.2005.}

\address{"T. Popoviciu" Institute of Numerical Analysis, Romanian
Academy, \\ P.O. Box 68-1, 400110 Cluj-Napoca, Romania}

\begin{abstract}
The log returns of financial time series are usually modeled by
means of the stationary GARCH(1,1)  stochastic process or its
generalizations which can not properly describe the nonstationary
deterministic components of the original series. We analyze the
influence of deterministic trends on the GARCH(1,1) parameters using
 Monte Carlo simulations. The statistical ensembles contain
numerically generated time series composed by GARCH(1,1) noise
superposed on deterministic trends. The GARCH(1,1) parameters
characteristic for financial time series longer than one year are
not affected by the detrending errors. We also show that if the ARCH
coefficient is greater than the GARCH coefficient, then the
estimated GARCH(1,1) parameters depend on the number of monotonic
parts of the trend and on the ratio between the trend and the noise
amplitudes.
\end{abstract}

\end{frontmatter}

\section{Introduction}

The log returns of financial time series $\{P_{t}\}$ (share prices,
stock indices, foreign exchange rates, etc.)
\begin{equation}
X_{t}=\log(P_{t}/P_{t-1})=\log\left(  P_{t}\right)  -\log\left(
P_{t-1}\right)  ~,\label{logret}%
\end{equation}
usually presents the following features: they are uncorrelated,
their volatility clusters, they have fat-tailed distributions
(leptokurtosis), a leverage effect is present (changes in stock
prices tend to be negatively correlated with changes in volatility),
their autocorrelation function decays exponentially, their absolute
values present a long range dependence \cite{mantStan}. One of the
most used stochastic models that reproduces some of these features
is the Generalized Auto Regressive Conditional Heteroskedasticity
(GARCH) model having the variance expressed as a linear function of
past squared innovations and earlier calculated conditional
variances \cite{boll86}. There are various generalizations of the
GARCH model, however the most used in practical applications is its
simplest form, GARCH(1,1).

According to relation (\ref{logret}) the stationary GARCH(1,1)
process is suitable for modeling time series for which $\{\log
P_{t}\}$ has a linear trend, i.e. the original time series
$\{P_{t}\}$ contains an exponential trend. But the nonlinear trends
in $\{\log P_{t}\}$ are not eliminated by the differentiation
(\ref{logret}). This problem is amplified in the case of a
nonmonotonic trend. An alternative to the stationary modeling of
financial series is the hypothesis of a nonstationary evolution. For
example St\u{a}ric\u{a} and Granger \cite{stagra2005} propose a
nonstationary model locally approximated  by a stationary one
\begin{equation}
X_{t}=\mu\left(  t\right)  +\sigma\left(  t\right) \varepsilon_{t}\
, \label{Xtnest}
\end{equation}
where $\varepsilon_{t}$ are i.i.d with $E(\varepsilon_{t})=0$ and
$E(\varepsilon_{t}^{2})=1$ and the unconditional mean $\mu\left(
t\right)  $ and the unconditional variance $\sigma\left(  t\right) $
are functions of $t.$ If a nonstationary series (\ref{Xtnest}) is
modeled with a  stationary process, then the deterministic trend
$\mu(t)$ is confounded with a stochastic trend and the model tends
to approach its nonstationarity limit. In the case of GARCH model
this is the so called IGARCH effect.

In this paper we study the influence of a deterministic trend on the
parameters of GARCH(1,1) model. We use a Monte Carlo method in order
to evaluate the influence of the detrending errors on the
variability of the estimated GARCH(1,1)\ parameters. The paper is
organized as follows. In the following section we shortly present
the GARCH(1,1) model and we study the intrinsic variability of its
parameters. In the third section an automatic method to generate
artificially trends is described. Then we present the variability of
the GARCH(1,1) parameters due to the detrending of an artificially
added deterministic component and the last section is dedicated to
conclusions.

\section{GARCH(1,1) model}

The GARCH(1,1) process is a real-valued discrete time stochastic
process
$\{x_{t}\}$%
\begin{eqnarray*}
&& x_{t}|\left\{  x_{t-1},\sigma_{t-1}\right\}     \sim  N\left(
0,\sigma
_{t}^{2}\right)  \\
&& \sigma_{t}^{2}  =  K+\alpha x_{t-1}^{2}+\beta\sigma_{t-1}^{2}%
\end{eqnarray*}
where $K>0,$ $\alpha\geq0,$ $\beta\geq0$ \cite{boll86}. If
$\alpha+\beta<1$, then GARCH(1,1) process is wide sense stationary
with $E\left(  x_{t}\right) =0,$ var$\left(  x_{t}\right)  =K/\left(
1-\alpha-\beta\right)  $ and cov$\left(  x_{t},x_{s}\right)
=$var$\left(  x_{t}\right)  \delta_{ts}$.

\begin{figure}
\begin{center}
\includegraphics{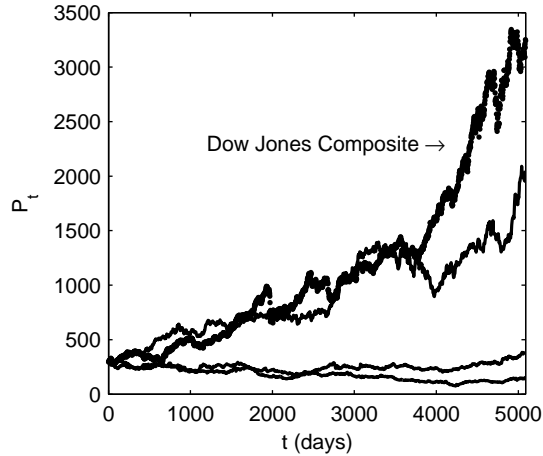}%
\caption{Dow Jones Composite index series over the period
01.02.1980-31.12.1999 and three index series simulated with GARCH(1,1) model}%
\label{figDJ1}
\end{center}
\end{figure}

We analyze the daily Dow Jones Composite (DJC) series containing
$N=5089$ values between 01 February 1980 and 31 December 1999 when a
deterministic trend is likely to exist. The parameters of the
GARCH(1,1) model obtained  using the maximum likelihood method for
the log returns of this series are: $\alpha_{DJ}=0.0837$,
$\beta_{DJ}=0.8898$, $K_{DJ}=2.5\cdot10^{-6}.$ With these values we
have generated three index series with the same initial value on the
same time interval. In Fig.~\ref{figDJ1} one observes large
differences between the generated series and the initial one,
especially for large $t$. This behavior is in accordance with the
fact that GARCH model is suitable only for relatively short periods
of time \cite{miksta}. When the series length is large, then the
GARCH parameters are close to the nonstationarity limit
($\alpha+\beta=1$) and the deterministic trend (if it exists) is
lost or is strongly distorted because it is replaced with a
stochastic one. In the case of the analyzed DJC index
$\alpha_{DJ}+\beta_{DJ}=0.9735$. Therefore, the variability of the
realizations of a GARCH process with given parameters is very large,
especially when the series length is large.

\begin{figure}
\includegraphics{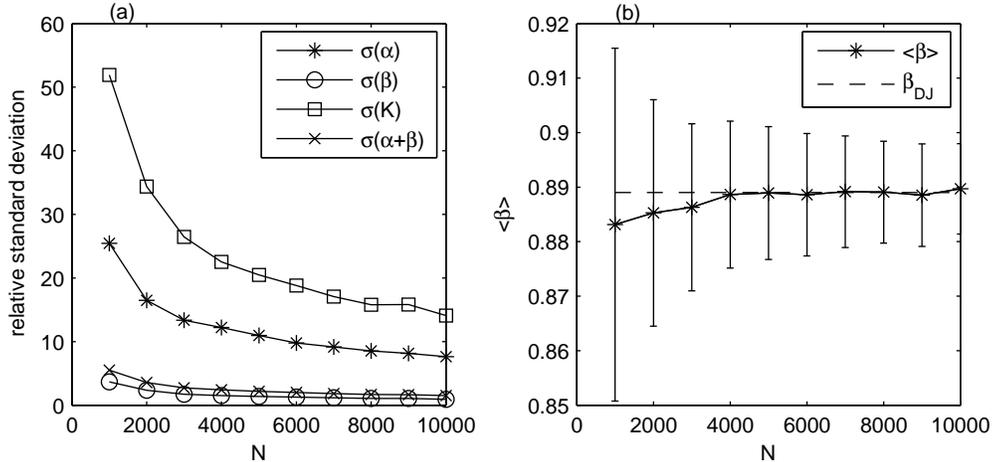}%
\caption{The relative standard deviation of the GARCH(1,1)
parameters (a) and the mean of $\beta$ (b) for statistical ensembles
with 500 numerically
generated series with DJ parameters for each length $N$.}%
\label{fig_var_Garch}%
\end{figure}

In order to correctly evaluate  the variability of the parameters
due to detrending, it must be compared with the intrinsic
variability of the GARCH(1,1) parameters for time series without
trends. The intrinsic variability is determined using a Monte Carlo
simulation. For a given length $N$ we generate 500 realizations of a
GARCH(1,1) process  with the parameters ($\alpha_{DJ}$,
$\beta_{DJ}$, $K_{DJ}$). For each realization the GARCH(1,1)
parameters are estimated by applying the maximum likelihood method.
The relative standard deviation of the GARCH(1,1) parameters
decreases to an almost stationary value for large values of $N$
(Fig.~\ref{fig_var_Garch}a). Generally, the mean of the estimated
parameters almost coincides with the values used for generating the
series with the exception of $\beta$ (Fig.~\ref{fig_var_Garch}b). In
the following we consider
series with length $N=6000$ which have a variability near the stationary value.%

\section{Automatically generated trends}

In order to be representative, the Monte Carlo statistics must
contain a large number of numerical simulations with variability
comparable with those appearing in practical applications. We
describe an automatic method for generating time series containing a
deterministic trend which satisfies these conditions. The generation
of a large number of trends with a significant variability using a
fixed functional form requires a large number of parameters. For
example, a polynomial trend must have a large enough degree, hence
the number of its coefficients is also large. If we choose the
coefficients by means of a random algorithm, then the form of the
generated trend is difficult to be controlled. Usually the resulting
trend has only a few parts with significant monotonic variation.

\begin{figure}
\includegraphics{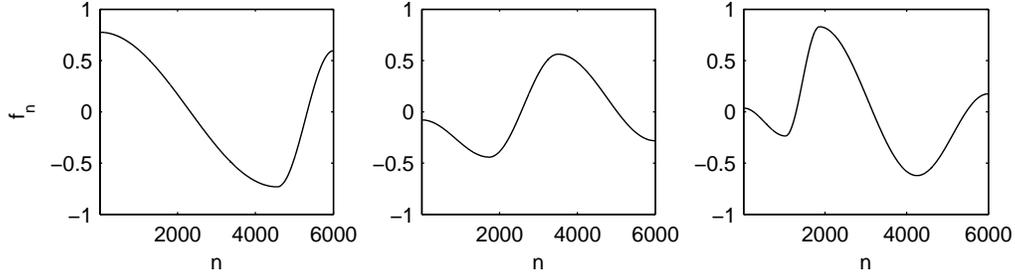}%
\caption{Artificially generated trends with 2, 3, and 4 monotonic parts}%
\label{tendinte}%
\end{figure}

We generate a trend $\left\{  f_{n}\right\}  $, $n=1,2,...,N$ by
joining together $s$ monotonic semiperiods of sinus with random
amplitudes and lengths. In this way we obtain a large enough
variability for the generated trends and we can control the number
and the amplitude of its monotonic parts. We need only three
parameters: the length of the series $N$ (in the following
$N=6000$), the number of monotonic parts $s$ (in our tests
$s=1,2,3,4$) and the minimum number of points in a part equal with
50. The amplitudes of the sinusoidal parts will be random numbers
with uniform distribution $a_{p}\in\left(  0,1\right)  $. The value
of the trend at the point $n$ of the
part $p$, $N_{p}<n\leqslant N_{p+1}$, is given by the recurrence relation%
\begin{equation}
f_{n}=f_{N_{p}}+(-1)^{p}a_{p}\left[  1-\sin\frac{\pi}{2}\left(
1+2\frac
{n-N_{p}}{N_{p+1}-N_{p}}\right)  \right]  ~,\label{fnprim}%
\end{equation}
where $f_{1}=0$. Some trends with different values for $s$ are
represented in
Fig.~\ref{tendinte}.%

We want to evaluate the error of the estimated GARCH(1,1) parameters
due to the difference between the estimated trend and the
real one. The statistical ensembles for the Monte Carlo simulations
are composed by numerically generated series composed by a random component
and the trend (\ref{fnprim}). First we generate a GARCH(1,1) time series
$\{x_{i}\}$ with given parameters ($\alpha_{0},\beta_{0},K_{0}$).
Then we calculate the series $y_{n}=\sum _{i=1}^{n}x_{i},$
(corresponding to the logarithm of a price series) and we add an
automatically generated trend, $\xi_{n}=f_{n}+y_{n}$. The relation
between these two components in the resulting series is
characterized by the
ratio $r$ of the amplitude of the trend and of the noise%
\[
r=\frac{\max(f_{n})-\min(f_{n})}{\max(y_{n})-\min(y_{n})}~.
\]
If we randomly choose the number $s$ of trend parts between two
given values $s_{\min}=1$ and $s_{\max}=4$ and the ratio $r$ between
$r_{\min}=0.25$ and $r_{\max}=4$ we obtain a significant statistics.

From the series $\{\xi_{n}\}$ we extract a polynomial estimated trend
$\{\widetilde{f}_{n}\}$, $\widetilde{y}_{n}=\xi_{n}-\widetilde{f}_{n},$
and we calculate the estimated returns $\widetilde{x}_{n}=\widetilde{y}%
_{n+1}-\widetilde{y}_{n}$. Then we evaluate the GARCH(1,1)
parameters ($\widetilde{\alpha},\widetilde{\beta},\widetilde{K}$)
using the maximum likelihood method. When we choose the degree of
the polynomial estimated trend we must take into account that if it
is too large, then the estimated trend begins to follow the
fluctuations of the noise. From numerical tests it has resulted an
optimal degree equal to $2s+3$.

\section{GARCH(1,1) parameters variability due to detrending}

\begin{figure}
\includegraphics{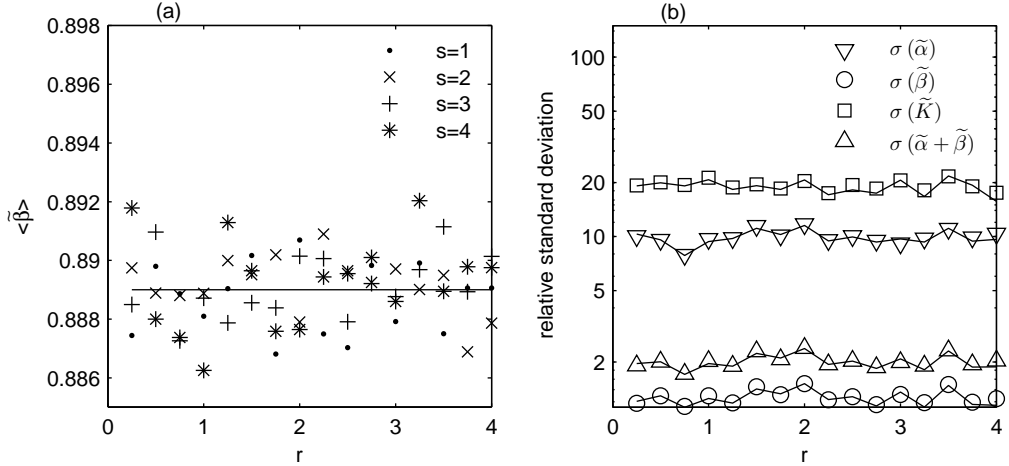}%
\caption{The variability of the GARCH(1,1) parameters due to
detrending errors with respect to the ratio $r$ between the trend and the noise
amplitude. The statistical ensembles contain 100 generated
GARCH(1,1) series with the DJC parameters superposed on
automatically generated trends. (a) The average value of the
estimated $\widetilde{\beta}$ for different numbers of trend
monotonic parts. The continuous line corresponds to the value
$\beta_{DJ}=0.8898$ used to generate the GARCH(1,1) noise. (b) The
relative standard deviation of the GARCH(1,1) parameters for the
generated noise $x_{n}$ (continuous line) and
for the estimated noise $\widetilde{x}_{n}$ (markers) for $s=4$.}%
\label{fig_vargarch_dj_trend}%
\end{figure}

Figure~\ref{fig_vargarch_dj_trend} shows the results obtained by
applying the evaluation method of the variability of GARCH(1,1)
parameters described in the previous section for statistical
ensembles of 100 time series generated with the DJC parameters for
different values of $r$ and $s$. From
Fig.~\ref{fig_vargarch_dj_trend}a it results that the averages of
the estimated parameter $\widetilde{\beta}$ are randomly distributed
around the initial value $\beta_{DJ}$ and they are not influenced by
the number of monotonic parts of the trend $s$ or by the ratio $r.$
The other GARCH(1,1) parameters have a similar behavior so we have
not represented them. Figure \ref{fig_vargarch_dj_trend}b confirms
this result by means of the relative standard deviation for $s=4$.
Hence, the GARCH(1,1) parameters are not influenced by detrending a
nonlinear trend if the noise is generated
using ($\alpha_{DJ}$, $\beta_{DJ}$, $K_{DJ}$).%

\begin{figure}
\includegraphics{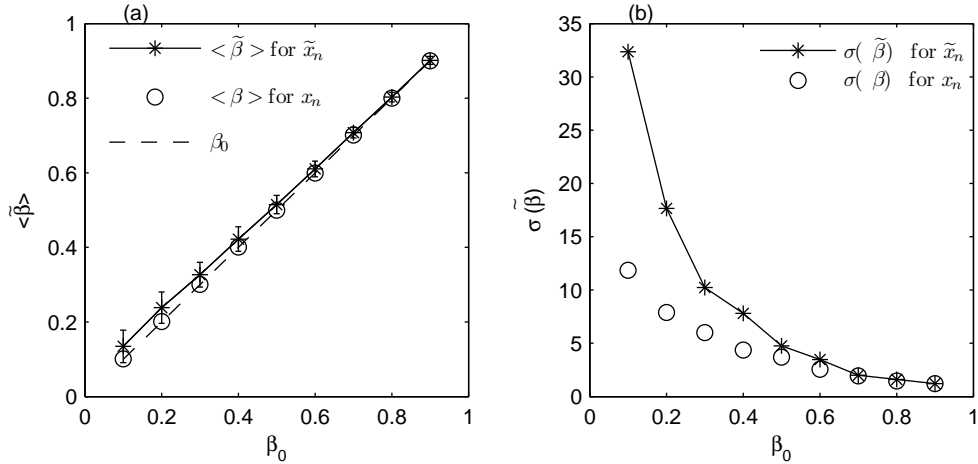}%
\caption{The variability of the parameter $\widetilde{\beta}$ due to
the detrending error for statistical ensembles of 100 generated
GARCH(1,1) series with the DJ parameters superposed on automatically
generated trends for $s=4$, $r=2$ and $\alpha_{0}+\beta_{0}=0.972$.
(a) The dashed line corresponds to the
values $\beta_{0}$ used to generate the GARCH(1,1) noise.}%
\label{med_dispg_gvect}%
\end{figure}

In our case the coefficient $\beta_{DJ}=0.8898$ has a much greater
value than $\alpha_{DJ}=0.083$. The variability of the GARCH(1,1)
parameters at detrending for greater ratios $\alpha_{0}/\beta_{0}$
is presented in Fig.~\ref{med_dispg_gvect}. The number of parts of
the artificially generated trend is $s=4$ and the ratio between
trend and noise is $r=2$ and $\alpha_{0}$ and $\beta_{0}$ are varied
such that $\alpha_{0}+\beta_{0}$ remains constant,
$\alpha_{0}+\beta_{0}=0.972$ and $K_0=K_{DJ}$. One observes that for
$\beta _{0}\geq 0.7$ the mean of the estimated values $\left\langle
\widetilde{\beta }\right\rangle $ almost coincides with the initial
value $\beta _{0}$ and the relative standard deviation is less than
3\%. Hence, the behavior observed for DJC index is the same for
smaller values of $\beta _{0}.\ $But the error $\left\langle
\widetilde{\beta}\right\rangle-\beta _{0}$
and the relative standard deviation $\sigma \left( \widetilde{%
\beta }\right) $ significantly increases when $\beta $ decreases
below $0.6$, so in this cases the influence of the errors of the
estimated trend is significant. The other two GARCH(1,1) parameters
have a similar behavior.

\begin{figure}
\includegraphics{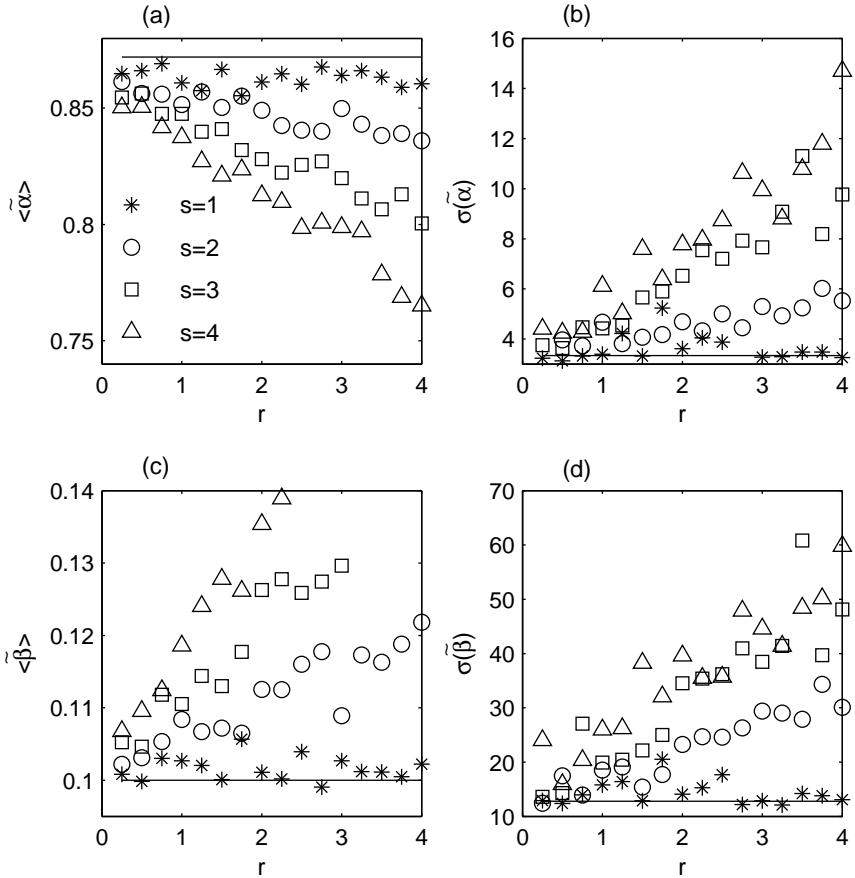}%
\caption{The mean and relative standard deviation of the GARCH(1,1)
parameters estimated for $\widetilde{x}_{n}$ obtained on statistical
ensembles of 100 time series for each $r$ and $s$. The continuous
line in (a) represents $\alpha_0$, in (c) $\beta_0$, in (b)
represents $\sigma(\alpha)$ and in (d) $\sigma(\beta)$.}%
\label{medstd_ag_a08g01}%
\end{figure}

Hence the influence of detrending on the variability of GARCH(1,1)
parameters is due especially to the coefficient $\beta$ that
generalizes the ARCH model. Figure~\ref{medstd_ag_a08g01} contains
the results of a detailed analysis for the minimum value
$\beta_0=0.1$ in Fig.~\ref{med_dispg_gvect}. The variability of the
estimated GARCH(1,1) parameters significantly increases when the
number $s$ of monotonic parts of the trend increases and the ratio
$r$ between the variation
amplitude of the trend and the noise is larger.%

\section{Conclusions}

An important problem in the analysis of financial series is to
separate the deterministic component and the stochastic one. In this
paper we have analyzed the influence of the nonstationarity due to a
deterministic trend on the GARCH(1,1) model and we have shown that
for long periods of tens of years the results obtained with this
model are not sensitive to the existence of a nonlinear trend. This
behavior occurs for large values of the GARCH parameter $\beta$ and
$\alpha+\beta$ close to 1. But for small values of $\beta$ \ and
large values of $\alpha$ such that $\alpha+\beta\simeq1$ the
influence of detrending on the GARCH(1,1) model becomes significant.

\end{document}